\documentclass[journal]{IEEEtran}

\ifCLASSINFOpdf
\usepackage[pdftex]{graphicx}
\graphicspath{{../pdf/}{../jpeg/}}
\DeclareGraphicsExtensions{.pdf,.jpeg,.png}
\else
\usepackage[dvips]{graphicx}
\graphicspath{{../eps/}}
\fi
\usepackage{graphicx}
\usepackage{epstopdf}
\usepackage{diagbox}
\usepackage{multirow}
\usepackage[cmex10]{amsmath}
\usepackage[ruled]{algorithm2e} 
\usepackage{url}
\usepackage{lineno,hyperref}
\usepackage{amsthm}
\usepackage{booktabs}
\usepackage{subfigure}
\usepackage{url}
\usepackage{mathrsfs}
\usepackage{amsmath}
\usepackage{amsthm}
\usepackage{bbding}

\usepackage{amsfonts,amssymb}
\usepackage{mathrsfs}
\usepackage{amsmath}
\usepackage{amsthm}
\usepackage{bbding}
\usepackage{cite}
\usepackage{xcolor}

\begin{document}
	\title{Semi-Supervised Disease Classification based on Limited Medical Image Data}
	\author{
		Yan Zhang, 
		Chun Li,
		Zhaoxia Liu,
		Ming Li
		\thanks{This research was supported by Guangdong Basic and Applied Basic Research Foundation (No.2021A1515220073).}
		\thanks{Y. Zhang and Z. Liu are with School of Science, Minzu University of China, Beijing, 100081, China. E-mail: za1234yuuy@gmail.com; liuzhaoxia@muc.edu.cn.}
		\thanks{C. Li is with Joint Research Center on Computational Mathematics and Control, Shenzhen MSU-BIT University, Shenzhen, 518172, China.}
		\thanks{M. Li is with Institute of Data Science, National University of Singapore, Singapore. E-mail: ming.li@u.nus.edu. }
		\thanks{*Corresponding author: Chun Li (E-mail: lichun2020@smbu.edu.cn).}
	}
	\markboth{IEEE Journal of Biomedical and Health Informatics}%
	{Shell \MakeLowercase{\textit{et al.}}: Bare Demo of IEEEtran.cls for IEEE Journals}
	\maketitle
	
	\begin{abstract}
		In recent years, significant progress has been made in the field of learning from positive and unlabeled examples (PU learning), particularly in the context of advancing image and text classification tasks. However, applying PU learning to semi-supervised disease classification remains a formidable challenge, primarily due to the limited availability of labeled medical images. In the realm of medical image-aided diagnosis algorithms, numerous theoretical and practical obstacles persist. The research on PU learning for medical image-assisted diagnosis holds substantial importance, as it aims to reduce the time spent by professional experts in classifying images. Unlike natural images, medical images are typically accompanied by a scarcity of annotated data, while an abundance of unlabeled cases exists. Addressing these challenges, this paper introduces a novel generative model inspired by Hölder divergence, specifically designed for semi-supervised disease classification using positive and unlabeled medical image data. In this paper, we present a comprehensive formulation of the problem and establish its theoretical feasibility through rigorous mathematical analysis. To evaluate the effectiveness of our proposed approach, we conduct extensive experiments on five benchmark datasets commonly used in PU medical learning: BreastMNIST, PneumoniaMNIST, BloodMNIST, OCTMNIST, and AMD. The experimental results clearly demonstrate the superiority of our method over existing approaches based on KL divergence. Notably, our approach achieves state-of-the-art performance on all five disease classification benchmarks.
		By addressing the limitations imposed by limited labeled data and harnessing the untapped potential of unlabeled medical images, our novel generative model presents a promising direction for enhancing semi-supervised disease classification in the field of medical image analysis. The code is available at the following URL: \url{https://github.com/wmh12138/Paper_Code/tree/master/train}.
	\end{abstract}
	
	\begin{IEEEkeywords}
		PU learning, generative model, medical image classification, divergence learning.
	\end{IEEEkeywords}
	
	\IEEEpeerreviewmaketitle

	\section{Introduction}
	\label{sec:1}
	\IEEEPARstart{P}ositive and unlabeled learning (PU learning) is a learning approach that aims to develop effective classifiers using only positive and unlabeled data, in contrast to traditional supervised learning (TSL) methods \cite{p1, p2, p3, p8,p54, p55}. PU learning poses unique challenges due to the limited availability of labeled data. To illustrate this point, let's consider the field of medical image processing, where the annotation of medical images requires the expertise of well-trained clinical professionals. However, these experts often face time constraints due to their busy schedules, making it challenging for them to manually review a large number of image samples. Consequently, the scarcity of labeled images hampers the development of effective classifiers.
	
	Deep learning models rely heavily on the availability of extensive training data and their corresponding expensive annotations. Unfortunately, the use of medical image-aided diagnosis is restricted by several factors \cite{p5, p6, p7, p80}:
	\begin{itemize}
		\item \textbf{Labor-intensive annotation}: The annotation of medical images is a labor-intensive task that demands considerable time and effort from clinical experts. Consequently, the lack of sufficient labeled data poses a fundamental challenge for medical image-aided diagnosis.
		\item \textbf{Exploitation of unlabeled data}: In addition to the limited number of labeled images, the collection of medical data often includes a significant amount of undiagnosed data, such as CT, MRI, and PET scans. These unlabeled samples represent a challenging problem as they lack corresponding labels and exhibit an uncertain illness state. Typically, these cases are considered ``suspicious''.
		\item \textbf{Expert involvement}: Medical image annotations necessitate the expertise of skilled clinicians, further impeding the availability of labeled data.
	\end{itemize}

Addressing these challenges requires innovative approaches that can effectively leverage both limited labeled data and the vast amount of unlabeled data. By doing so, PU learning methods hold great potential for enhancing the accuracy and efficiency of medical image analysis and aiding in diagnosis.

PU learning is a valuable approach in the realm of semi-supervised binary classification, specifically designed to handle ambiguous data scenarios \cite{p8, p10, p11}. It aims to recover labels from unknown samples by leveraging patterns learned from positive cases. PU learning has gained significant attention as a prominent area of research in semi-supervised learning, particularly applicable to disease classification based on medical image analysis. Various categories of PU learning methods have emerged, including:
\begin{itemize}
	\item \textbf{Two-stage method} \cite{p14, p15}: This approach involves two steps. Firstly, reliable negative examples are extracted from the unlabeled sample set. In the second step, a supervised or semi-supervised learning algorithm is employed to train the classifier. However, this method requires the utilization of information from the positive example set to compile an initial collection of trustworthy negative examples from the unlabeled set.
	
	\item \textbf{Statistical learning based on positive samples}: Denis et al. \cite{p16} proposed a naive Bayesian classifier using the statistical query learning model, tailored to PU learning. However, these models often necessitate the provision of prior probability for positive examples.
	
	\item \textbf{One-sided noise learning}: This approach treats unlabeled samples as noisy negative examples, reducing the PU learning problem to an one-sided noise learning problem. Examples include biased SVM \cite{p18} and biased PRTFIDF \cite{p19}. However, overfitting remains a challenge. To address this, Kiryo et al. \cite{p17} proposed non-negative PU learning, improving the objective function based on the unbiased learning model.
	
	\item \textbf{Predictive adversarial networks (PANs)}: Introduced by Hu et al. \cite{p20}, this method introduces a new objective function based on KL divergence and generative adversarial networks (GANs). The generator $G$ of GANs is replaced by a classifier $C$, and the optimization goal is to train a discriminator $D$ for classifying unlabeled data. The PAN learning approach has been shown, both theoretically and experimentally, to train a binary classifier without relying on prior probability. It outperforms current state-of-the-art PU learning models such as NNPU \cite{p17} and UPU \cite{p21}.
\end{itemize}

Previous works in this field can be categorized into three main groups \cite{p20, p34, p35, p36, p37, p38, p39}. However, each of these categories has its limitations. For instance, the two-stage method requires utilizing information from the positive example set in the initial stage to derive a reliable negative example set from the unlabeled samples. In statistical learning based on positive samples, users need to provide the prior probability of positive examples, and overfitting issues often necessitate the application of counterexample bias methods. In contrast, the proposed algorithm effectively circumvents these aforementioned challenges.

To address the limitations of previous approaches, we propose and evaluate a novel HD-PAN based on Hölder divergence, inspired by its appealing mathematical properties. We investigate the influence of the objective function, exploring different divergence formulas to enhance classification results. Hölder divergence is found to play a crucial role in PU learning-based classification. Our method is extensively evaluated on several medical image datasets, demonstrating its effectiveness and robustness. We achieve state-of-the-art performance on five disease classification benchmarks, namely BreastMNIST, PneumoniaMNIST, BloodMNIST, OCTMNIST \cite{p25, p26}, and AMD.

In summary, our contributions in this work can be outlined as follows:
\begin{enumerate}
	\item Proposal of the HD-PAN, leveraging the mathematical properties of Hölder divergence.
	
	\item Investigation of the objective function, introducing improvements based on different divergence formulas, highlighting the significance of Hölder divergence in PU learning-based classification.
	
	\item Attainment of state-of-the-art performance on five disease classification benchmarks, BreastMNIST, PneumoniaMNIST, BloodMNIST, OCTMNIST \cite{p25, p26}, and AMD.
\end{enumerate}

In this work, we introduce several notations that are used throughout the work. These notations are summarized in Table \ref{tab-1}.

\begin{table}[t]
	\centering
	\caption{Main Notations Used in This Work.}
	\setlength{\tabcolsep}{3pt}
	\begin{tabular}{p{1.8cm}<{\raggedright}|p{6cm}<{\raggedright}} 
		\toprule [1.5pt]
		Notation&Definition\\
		\midrule [0.5pt]
		$p(z)$& noise distribution \\
		$G(z)$ & generated samples by $G(.)$ \\
		$x \sim {p_{data}}(x)$ & real sample\\
		$P^p$ & generative distribution of labeled data in PU learning \\
		$P^u$ & generative distribution of unlabeled data in PU learning \\
		$X^p, X^u$ & positive and unlabeled dataset, respectively\\
		$X^{pu}$& given PU dataset\\
		$P_i^{pu}$& given PU data $X^{pu}$ the true probability distribution of the positive and unlabeled two results of the $i$th sample\\
		\bottomrule[1.5pt]
	\end{tabular}
	\label{tab-1}
\end{table}	
\section{Related Works}
\label{sec:2}
\subsection{Generative Adversarial Networks}
GANs, initially proposed by Ian et al. in 2014 \cite{p27, p28}, are adversarial learners designed to train a generator that can produce data resembling real samples. GANs consist of two main components: the discriminator $D$ and the generator $G$. In essence, the generator $G$ aims to deceive the discriminator $D$ by generating synthetic samples, while the discriminator $D$ strives to differentiate between the generated samples $G(z)$ and the true data samples from $p_{data}(x)$. The objective functions for training GANs are as follows:
\begin{equation}
	\label{equ_1}
	\left\{ {\begin{array}{*{20}{c}}
			{\mathop {\max L(D) = }\limits_D \left( \begin{array}{l}
					{E_{x \sim {p_{data}}(x)}}\left[ {\log D(x)} \right]\\
					+ {E_{z \sim {p_z}(z)}}\left[ {\log (1 - D(G(z)))} \right]
				\end{array} \right),}\\
			{\mathop {\min L\left( G \right)}\limits_G  = {E_{z \sim {p_z}(z)}}\left[ {\log (1 - D(G(z)))} \right].\begin{array}{*{20}{c}}
					{}&{}&{}
			\end{array}}
	\end{array}} \right.
\end{equation}

GANs possess several advantages. Firstly, they require relatively low computational resources compared to other generative models. Additionally, GANs are capable of generating high-quality images. Moreover, GANs employ an unsupervised learning approach, making them applicable in unsupervised and semi-supervised learning domains. These advantages have led to the widespread adoption of GANs in medical image analysis and computer vision. For instance, Yu et al. \cite{p29} employed GANs to synthesize cross-modality MR images by establishing global and local mappings. Upadhyay et al. \cite{p30} utilized uncertainty-guided progressive GANs for medical image translation. Marouf et al. \cite{p31} proposed a conditional single-cell CSCGAN to generate realistic single-cell RNA-seq data. Cen and David \cite{p32} improved protein function prediction by employing synthetic feature samples and GANs. Jiang et al. \cite{p70} employed a conditional GAN to synthesize COVID-19 CT images. Yoon et al. \cite{p71} introduced the ADS-GAN framework for data synthesis. Zhao et al. \cite{p72} utilized a GAN for the prediction of Alzheimer's disease progression. Hu et al. \cite{p73} developed unsupervised learning methods based on GAN for cell-level visual representation. Chen et al. \cite{p74} proposed JAS-GAN for the segmentation of unbalanced data. Phattarapong et al. \cite{p75} developed EEGANet to remove ocular artifacts from EEG signals. Zhou et al. \cite{p76} developed DR-GAN for fine-grained lesion synthesis on diabetic retinopathy images. Luo et al. \cite{p77} employed a GAN for dehazing cataractous retinal images. Yang et al. \cite{p78} developed a semi-supervised approach based on GAN for bi-modality medical image synthesis. For further exploration of GANs, refer to \cite{p60, p61, p62}, which provide more theoretical and applied research on GANs.

\subsection{Positive and Unlabeled Learning}
Currently, PU learning has gained significant attention in the field of semi-supervised learning, and researchers have made notable contributions in this area. PU learning is a binary classification algorithm that assigns only two values, positive and negative to the data. PU learning leverages a small amount of labeled positive data in the training phase to infer the characteristics of positive samples and then applies these learned characteristics to classify unknown data and recover their labels.

For instance, Xu et al. \cite{p34} developed a one-step strategy to directly train a multi-class classifier using given multi-class data. Gong et al. \cite{p35} employed traditional decision margin-based methods to train classifiers, but the model performance was subpar, particularly in image classification scenarios. Bekker and Davis \cite{p36} explored random assumptions for PU learning, while Bao et al. \cite{p37} investigated multi-component PU learning models. Sansone et al. \cite{p38} focused on scalable PU learning models. Unfortunately, these models often require prior probability estimation, which can be a challenge. Zhao et al. \cite{p39} utilized entropy minimization principles and mixup regularization to address the issue of trivial label distribution consistency. However, it has been observed that incorrect estimation of the prior probability can lead to poor results. Addressing these limitations, Hu et al. \cite{p20} proposed an adversarial training network-based classification model, integrating the adversarial principle into PU learning for the first time. Their approach does not rely on prior knowledge and demonstrates excellent performance in terms of classification accuracy, sensitivity, and robustness.

\section{Background}
\label{sec:3}
\subsection{Hölder Divergence}
Kullback-Leibler (KL) divergence is a metric used to quantify the similarity between two probability distributions. For a random variable $x$, the KL divergence from distribution $P$ to distribution $Q$ is defined as:

\begin{equation}
	\label{equ_2}
	{D_{KL}}(P||Q) = \int {p(x)\ln \left( {\frac{{p(x)}}{{q(x)}}} \right)dx},
\end{equation}

In 2014, the concept of ``Hölder divergence'' was introduced, building upon the definition of the Hölder score \cite{p22, p23, p24}. It utilizes the tightness of inequalities to establish a dissimilarity measurement system. For an inequality expressed as $lhs \le rhs$, where $lhs$ and $rhs$ denote the left and right sides of the inequality respectively, the tightness, denoted as $\Delta = rhs - lhs$, is evaluated. When $lhs > 0$, the following difference can be defined:

\begin{equation}
	\label{equ_3}
	D = \log \frac{{rhs}}{{lhs}} = - \log \frac{{lhs}}{{rhs}} \ge 0,
\end{equation}

For a two-parameter inequality $lhs(p,q) \le rhs(p,q)$ where $p \ne q$, if $lhs(p,q) < rhs(p,q)$ is always true, the inequality is called proper. If $p=q$ and $lhs(p,q) = rhs(p,q)$ holds, then the inequality is considered tight. Furthermore, a proper inequality can define a proper difference; that is, $q=p$ and $D(p,q) = 0$ represent a proper fix. Inequalities other than proper inequalities are referred to as ill-posed inequalities. Therefore, pseudo-divergence is established based on ill-posed inequalities. Hölder divergence is defined using the Hölder inequality, which states that for two positive real-valued functions $p(x)$ and $q(x)$ in the same probability space $\Omega$, the following relationship holds:

\begin{equation}
	\label{equ_4}
	\int_\Omega {p(x)q(x)dx} \le {\left( {\int_\Omega {p{{(x)}^\alpha }dx} } \right)^{\frac{1}{\alpha }}}\left( {\int_\Omega {q{{(x)}^\beta }dx} } \right)^{^{\frac{1}{\beta }}},
\end{equation}

where $\alpha$ and $\beta$ satisfy $\alpha\beta>0$ and $\frac{1}{\alpha } + \frac{1}{\beta } = 1$. They are referred to as a pair of Hölder conjugate exponents. For the Hölder conjugate exponents $\alpha$ and $\beta$, if the probability density functions $p(x) \in {L^\alpha }(\Omega ,\mu )$ and $q(x) \in {L^\beta}(\Omega ,\mu )$, then the Hölder pseudo-divergence is defined as:

\begin{equation}
	\label{equ_5}
	D_\alpha ^H\left( {p(x):q(x)} \right) = - \log \left( {\frac{{\int_\Omega {p(x)q(x)dx} }}{{{{\left( {\int_\Omega {p{{(x)}^\alpha }dx} } \right)}^{\frac{1}{\alpha }}}{{\left( {\int_\Omega {q{{(x)}^\beta }dx} } \right)}^{\frac{1}{\beta }}}}}} \right),
\end{equation}

where $\mu$ and $L(\Omega ,\mu )$ refer to the Lebesgue measure and Lebesgue measure space, respectively.
	\begin{figure*}[t]
	\centering
	\includegraphics[width=1.1\linewidth]{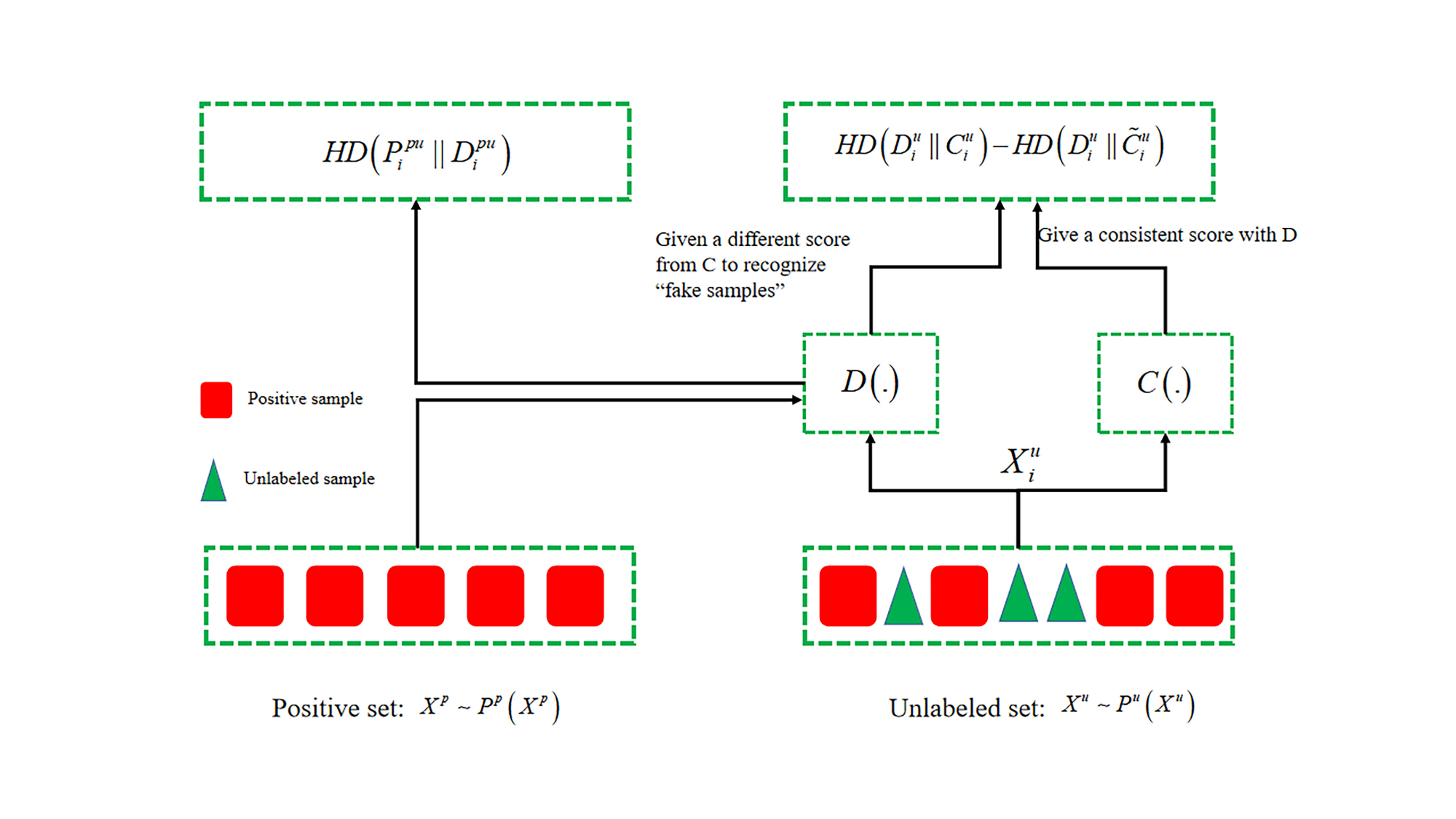}
	
	\caption{The Framework of HD-PAN. HD-PAN presents a comparison between two models, $D(\cdot)$ and $C(\cdot)$. The classifier $C(\cdot)$ aims to assess the probability of the unlabeled data from set $U$ being positive data, subsequently forwarding the data to the discriminator $D(\cdot)$ to determine its authenticity as positive data. }
	\label{fig_1}
\end{figure*}

\section{Methodology}
\label{sec:4}

This section introduces a semi-supervised disease classification strategy based on Hölder divergence, referred to as HD-PAN for brevity. PU learning is employed as the underlying semi-supervised approach. 

\subsection{Motivation}	
The primary objective of this study is to classify diseases using medical image databases. Therefore, it is necessary to learn the features present in medical images and train a classifier accordingly. However, diseased features in medical images are not limited to a single location, and images depicting diseases often exhibit diseased characteristics in multiple regions. To achieve the goal of accurately detecting lesion areas within images, the deep learning algorithm must effectively identify these affected regions, making it a multi-classification problem. In the domain of class distribution recognition research, KL divergence demonstrates a significant increase, reaching its maximum value at the boundary. Consequently, in multi-class recognition, KL divergence tends to focus on a few categories with more prominent properties, disregarding others. In contrast, Hölder divergence exhibits distinct characteristics, presenting a gradually increasing curve and the ability to recognize all groups, including those with evident traits. In this experiment, Hölder divergence is employed as a substitute for KL divergence.

Furthermore, Hölder divergences encompass both the Cauchy-Schwarz and the one-parameter family of skew Bhattacharyya divergences, making them noteworthy. In cases where the natural parameter space is a cone or an affine, Hölder divergences allow for closed-form expressions between distributions belonging to the same exponential families. This is particularly advantageous when studying exponential families with conic or affine natural parameter spaces, such as multinomials or multivariate normals. Both statistical Hölder distance families enable the use of closed-form formulas and do not necessitate distribution normalization, as they are inherently appropriate divergences between probability densities \cite{p23, p24}.

\subsection{Problem Statement}
It is well-known that GANs consist of a generator $G(\cdot)$ and a discriminator $D(\cdot)$. However, in PU learning, the primary goal is to train a predictor or classifier that accurately predicts data labels, denoted as $C(\cdot)$. Instead of training a generator $G(\cdot)$, the classifier $C(\cdot)$ aims to determine the probability of unlabeled data from set $U$ being positive data. Subsequently, the data is passed to the discriminator $D(\cdot)$ to determine its authenticity as positive data. In our HD-PAN, we retain the discriminator $D(\cdot)$ but replace the generator $G(\cdot)$ with the classifier $C(\cdot)$ to achieve the classification objective. Mathematically, this can be described by the following formulation:

\begin{equation}
	\label{equ_5}
	\begin{array}{l}
		\mathop {\min }\limits_C \mathop {\max }\limits_D V(D,C)\begin{array}{*{20}{c}}
			{}&{}&{}&{}&{}&{}&{}&{}
		\end{array}\\
		\begin{array}{*{20}{c}}
			{}&{}
		\end{array} = \left( \begin{array}{l}
			{E_{{x^p} \sim {P^p}({x^p})}}\left[ {\log D({x^p})} \right]\\
			+ {E_{{x^s} = {{\arg }_{{x^u} \sim {P^u}({x^u})}}C({x^u}) = 1}}\left[ {\log (1 - D({x^s}))} \right]
		\end{array} \right),
	\end{array}
\end{equation}
where $x_s$ represents data from the unlabeled data set $U$ that is classified as positive by the classifier $C(\cdot)$. $P^p$ denotes the generation distribution of labeled data in PU learning, and $P^u$ represents the generative distribution of unlabeled data in PU learning. As the last term in the equation is discrete, the model's inventors trained the model using policy gradients from reinforcement learning. The framework of HD-PAN is illustrated in Figure \ref{fig_1}.

\subsection{Objective Function}	
The objective of PU learning is to train a classifier $C(\cdot)$ to classify unlabeled data. Despite incorporating the idea of adversarial networks, training a classifier $C(\cdot)$ remains essential. Therefore, we can consider the classifier $C(\cdot)$ and the discriminator $D(\cdot)$ as two probability distributions. The objective of the classifier $C(\cdot)$ is to fit the entire probability distribution of the discriminator $D(\cdot)$, while the discriminator $D(\cdot)$ aims to differentiate itself from the probability distribution of the classifier $C(\cdot)$. The approach can be summarized as follows:
\begin{enumerate}
	\item The classifier $C(\cdot)$ assigns high probabilities to data that is challenging for the discriminator $D(\cdot)$ to distinguish and low probabilities to data that the discriminator $D(\cdot)$ can easily differentiate.
	
	\item The discriminator $D(\cdot)$ needs to discriminate between samples classified as positive data by the classifier $C(\cdot)$ and the real positive data. The probability here refers to the likelihood of the data being positive.
\end{enumerate}	

Additionally, the original PAN assumes that the output results of samples after passing through $D(\cdot)$ and $C(\cdot)$ follow a 0-1 distribution. Considering the analysis above and the mathematical properties of Hölder divergence compared to KL divergence, it is more reasonable to use Hölder divergence as a measure of the distance between two probability distributions. This leads to the following objective function of HD-PAN:
\begin{equation}
	\label{equ_7}
	\begin{array}{l}
		\mathop {\min }\limits_C \mathop {\max }\limits_D V(D,C)\begin{array}{*{20}{c}}
			{}&{}&{}&{}&{}&{}&{}&{}
		\end{array}\begin{array}{*{20}{c}}
			{}&{}&{}&{}&{}&{}&{}&{}
		\end{array}\\
		\begin{array}{*{20}{c}}
			{}&{}&{}
		\end{array} =  - \sum\limits_{i = 1}^n {D_\alpha ^H(P_i^{pu}||D_i^{pu})} \\
		\begin{array}{*{20}{c}}
			{}&{}&{}
		\end{array} + \lambda (\sum\limits_{i = 1}^{{n_0}} {D_\alpha ^H(D_i^u||C_i^u) - } \sum\limits_{i = 1}^{{n_0}} {D_\alpha ^H(D_i^u||\hat C_i^u)} ),
	\end{array}
\end{equation}
where $P_i^{pu}$ represents the positive sample of the $i$-th sample in the given PU data $X^{pu}$ (including the positive data set $X^p$ and the unlabeled set $X^u$) and the true probability distribution of the two unknown results. Here, $n$ is the total number of data in the sample set $X^{pu}$ during training, and $n_0$ is the total number of data in the unlabeled data set $X^u$ during training. $\hat C_i^u$ represents the inverse distribution of $C_i^u$, and $\lambda$ is a hyperparameter for balancing the distance. By replacing $D(x^{pu})$ and $C(x^{pu})$ with $D$ and $C$, respectively, the objective function can be simplified as follows:
\begin{equation}
	\label{equ_8}
	\begin{array}{l}
		\mathop {\min }\limits_C \mathop {\max }\limits_D V(D,C)\begin{array}{*{20}{c}}
			{}&{}&{}&{}&{}&{}&{}&{}
		\end{array}\\
		\begin{array}{*{20}{c}}
			{}&{}&{}
		\end{array} = \left( \begin{array}{l}
			{E_{{x^p} \sim {P^p}({x^p})}}[\log D({x^p})]\\
			+ {E_{_{{x^u} \sim {P^u}({x^u})}}}[\log (1 - D({x^u}))]
		\end{array} \right)\\
		\begin{array}{*{20}{c}}
			{}&{}&{}
		\end{array} + \left( \begin{array}{l}
			\lambda {E_{_{{x^u} \sim {P^u}({x^u})}}}[(\log (1 - C({x^u}))\\
			- \log (C({x^u})))(2D({x^u}) - 1)]
		\end{array} \right),
	\end{array}
\end{equation}	
where $P^p$ represents the distribution of positive data. The algorithmic flowchart is depicted in Algorithm \ref{algorithm_1}.

	\begin{algorithm}[!t]
	\caption{\small Training HD-PAN.}
	\label{algorithm_1}
	\DontPrintSemicolon
	\small
	\textbf{Input:} Dataset $\mathcal{D}$, positive training data $X^p$ and unlabeled training data $X^u$;  Randomly initialized HD-PAN; $t=0$,
	
	$D(.)$, $C(.)$, $\leftarrow$ Randomly initialize network parameters
	
	\While{number of training steps} 
	{ \tcp*[f]{Training $D(.)$ $k$ steps, we set $k = 1$.} \\
			\While{k steps}  
		   { 
			Sample a mini-batch of $m$ positive examples $\left\{ {x_1^p, \ldots ,x_m^p} \right\}$ from $X^p$\\
			Sample a mini-batch of $m$ unlabeled examples $\left\{ {x_1^u, \ldots ,x_m^u} \right\}$ from $X^u$\\
			Update $D(.)$ by ascending its stochastic gradient:
		    $${\nabla _{{\Theta _d}}}\left( \begin{array}{l}
		    	- \sum\limits_{i = 1}^n {D_\alpha ^H(P_i^{pu}||D_i^{pu})} \\
		    	+ \lambda (\sum\limits_{i = 1}^{{n_0}} {D_\alpha ^H(D_i^u||C_i^u)} \\
		    	- \sum\limits_{i = 1}^{{n_0}} {D_\alpha ^H(D_i^u||\hat C_i^u)} )
		    \end{array} \right)$$
			
		   }
	     Sample a mini-batch of $m$ unlabeled examples $\left\{ {x_1^u, \ldots ,x_m^u} \right\}$ from $X^u$\\
	     Update $C(.)$ by ascending its stochastic gradient:
	     $${\nabla _{{\Theta _c}}}\left( \begin{array}{l}
	     	- \sum\limits_{i = 1}^n {D_\alpha ^H(P_i^{pu}||D_i^{pu})} \\
	     	+ \lambda (\sum\limits_{i = 1}^{{n_0}} {D_\alpha ^H(D_i^u||C_i^u)} \\
	     	- \sum\limits_{i = 1}^{{n_0}} {D_\alpha ^H(D_i^u||\hat C_i^u)} )
	     \end{array} \right)$$
	}
\end{algorithm}

\section{Experiments and Analysis}
\subsection{Datasets}	
\begin{figure} 
	\centering
	\includegraphics[width=0.7\linewidth]{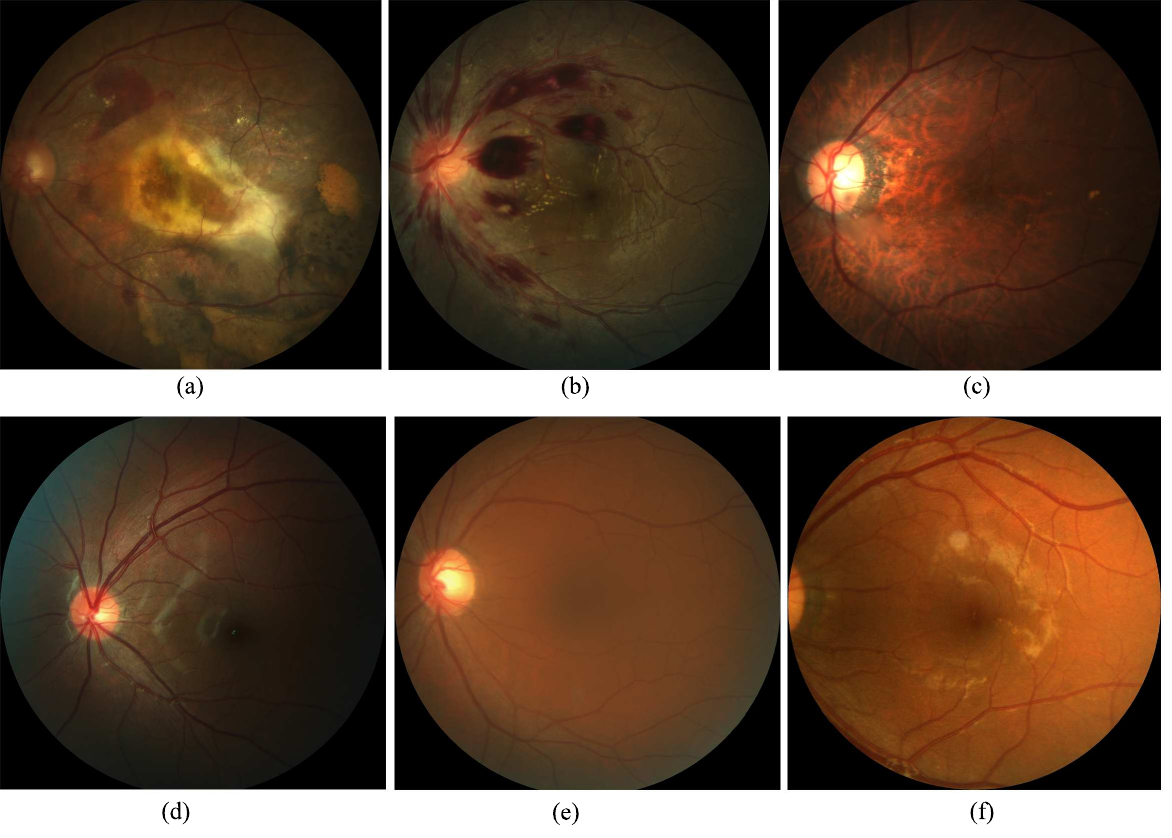}
	
	\caption{AMD and Non-AMD Examples: Panels (a)-(c) showcase examples of AMD, while panels (d)-(f) depict examples of Non-AMD. }
	\label{fig_2-add}
\end{figure}

To enhance the model's logical coherence, result reliability, and algorithm robustness for disease classification, this study employs the MedMNIST \cite{p25, p26}, a widely recognized medical imaging dataset, for testing purposes, and Age-related Macular degeneration (AMD) (\url{https://amd.grand-challenge.org/}). MedMNIST, developed and made available by researchers from Shanghai Jiaotong University, encompasses ten sub-datasets consisting of diverse medical imaging modalities. In this work, the focus is on five specific sub-datasets: BreastMNIST, PneumoniaMNIST, OCTMNIST, BloodMNIST, and AMD. A brief overview of these datasets is provided below:
\begin{itemize}
	\item \textbf{BreastMNIST} \cite{p25, p26}: This dataset comprises two classes and consists of 546 training images, 78 validation images, and 156 test images, all labeled accordingly. Each image has a size of $28 \times 28$ pixels.
	
	\item \textbf{PneumoniaMNIST} \cite{p25, p26}: With a binary classification task, the PneumoniaMNIST dataset contains 4708 training images, 524 validation images, and 624 test images, accompanied by respective labels. Each image in this dataset is of size $28 \times 28$ pixels.
	
	\item \textbf{OCTMNIST} \cite{p25, p26}: This dataset encompasses five classes and includes 97,477 training images, 10,832 validation images, and 1000 test images, each with corresponding labels. The images in OCTMNIST are of size $28 \times 28$ pixels.
	
	\item \textbf{BloodMNIST} \cite{p25, p26}: With eight classes, the BloodMNIST dataset consists of 11,959 training images, 1712 validation images, and 3421 test images, all labeled accordingly. The images in this dataset are also sized $28 \times 28$ pixels.
	\item \textbf{Age-related Macular degeneration (AMD)}\cite{p79}: The Age-related Macular Degeneration (AMD) dataset serves as the foundation for evaluating sophisticated algorithms designed for the diagnosis of AMD, along with the segmentation and classification of lesions in fundus photos obtained from AMD patients. The primary aim is to assess and compare automated techniques for detecting AMD, leveraging a standardized dataset of retinal fundus images. This classification task focuses on a binary dataset. The training set encompasses 240 images, each paired with corresponding labels, while the test set includes 160 labeled images. The images are formatted in colored RGB, with dimensions measuring $2124 \times 2056$ pixels. AMD and Non-AMD examples are shown in Figure \ref {fig_2-add}.
\end{itemize}

By utilizing these diverse and representative sub-datasets from MedMNIST and AMD, this research aims to achieve improved disease classification performance while ensuring the validity and generalizability of the findings.

\subsection{Training Details}
\subsubsection{Evaluation Metrics}
To effectively evaluate the performance of our disease classification method, we choose accuracy, recall, precision, and $F_{1}$-score as our evaluation metrics \cite{p43}. These metrics provide comprehensive insights into the model's performance and are defined as follows:
\begin{itemize}
	\item \textbf{Accuracy}: Accuracy measures the overall correctness of the classification and is computed as the ratio of the sum of true positives (TP) and true negatives (TN) to the total number of samples: $Accuracy = \frac{{TP + TN}}{{TP + TN + FP + FN}}$.
	
	\item \textbf{Recall}: Recall, also known as sensitivity or true positive rate, measures the ability of the classifier to correctly identify positive instances. It is calculated as the ratio of TP to the sum of TP and false negatives (FN): ${\rm{Recall}} = \frac{{TP}}{{TP + FN}}$.
	
	\item \textbf{Precision}: Precision quantifies the accuracy of positive predictions made by the classifier. It is calculated as the ratio of TP to the sum of TP and false positives (FP): ${\rm{Precision}} = \frac{{TP}}{{TP + FP}}$.
	
	\item \textbf{$F_{1}$-score}: The $F_{1}$-score is the harmonic mean of precision and recall, providing a balanced measure that combines both metrics. It is calculated as: ${F_1}\text{-score} = 2 \times \frac{{\text{{Precision}} \times \text{{Recall}}}}{{\text{{Precision}} + \text{{Recall}}}}$.
\end{itemize}

By considering these evaluation metrics, including accuracy, recall, precision, and $F_{1}$-score, we can comprehensively assess the performance of our disease classification method. These metrics capture different aspects of the model's performance, such as overall accuracy, the ability to detect positive instances, and the balance between precision and recall.

\subsubsection{The Experiment's Purpose}
Our experimental study aims to achieve the following three objectives:
\begin{itemize}
	\item \textbf{Improving the objective function of PAN}: In this work, we propose an enhanced objective function for PAN, known as HD-PAN. By refining the objective function and adjusting the Hölder parameters, our goal is to investigate whether HD-PAN can outperform the original PAN in terms of disease classification accuracy and recall. We aim to demonstrate the effectiveness of the improved objective function in enhancing the performance of disease classification.
	
	\item \textbf{Comparing HD-PAN with state-of-the-art models}: We compare HD-PAN with other state-of-the-art models. To ensure a fair comparison, we utilize the same batch of datasets, network structures, and training parameters for all models. By training these models on the identical datasets, we can evaluate their performance based on accuracy and $F_1$-score. This comparative analysis will help determine which model achieves better classification results and more accurate predictions.
	
	\item \textbf{Evaluating HD-PAN for disease classification}: We aim to explore the accuracy and effectiveness of HD-PAN in disease classification tasks. By conducting experiments on five medical imaging datasets, we utilize accuracy and $F_1$-score as evaluation metrics to assess the performance of HD-PAN. These metrics will enable us to measure the accuracy of disease classification and gauge the overall effectiveness of HD-PAN in comparison to other models.
\end{itemize}

By addressing these three goals, our experiment aims to contribute to the advancement of disease classification algorithms, enhance the understanding of HD-PAN's capabilities, and provide insights into its potential applications in medical image processing.
\subsection{Preparation for Experiment}
\subsubsection{Data Preprocessing}
To conduct PU learning, it is essential to have a dataset comprising both labeled and unlabeled samples. However, the number of labeled samples is often limited compared to the total dataset size. Furthermore, the algorithm's ultimate objective is to achieve binary classification results, distinguishing between positive and negative classes. Consequently, preprocessing of the five medical imaging datasets is carried out prior to the experiment.

For BreastMNIST and PneumoniaMNIST, the datasets already consist of two distinct types and therefore require no additional processing. For the AMD dataset, we randomly select 50 images to form the positive dataset.

OCTMNIST, on the other hand, is a four-category dataset with category labels ranging from 1 to 4. In order to convert it into a binary classification task, the data is divided into positive and negative samples based on the parity of the category labels.

BloodMNIST, an eight-category dataset with category labels from 1 to 8, also required preprocessing. The data is divided into positive and negative samples based on the parity of the category labels, designating images with even-numbered categories as positive samples and those with odd-numbered categories as negative samples.

By performing this preprocessing step, we ensure that all five medical imaging datasets are suitably prepared for the PU learning experiment. The resulting datasets allowed for effective training and evaluation of the proposed method's performance in disease classification tasks.

\subsubsection{The Establishment of the PU Learning Data Set}
To construct the training dataset for PU learning, we obtain positive and negative training samples for each dataset. For datasets with a limited number of samples, namely BreastMNIST and PneumoniaMNIST, we randomly select 100 images as the positive data set. This selection ensure a representative subset of positive samples for training.

In contrast, for datasets with a larger volume of data, namely OCTMNIST and BloodMNIST, we aim to maintain a suitable balance between positive and negative samples. We randomly select 2000 images as the positive data set for OCTMNIST and 500 images as the positive data set for BloodMNIST. These selections are made to ensure an appropriate distribution of positive samples within the overall training dataset.

By constructing the PU learning training dataset in this manner, we ensure a combination of labeled and unlabeled data from each dataset. This dataset composition facilitates effective training of the proposed method and allows for accurate disease classification in subsequent experiments.

\begin{figure*}
	\centering
	\includegraphics[width=0.9\linewidth]{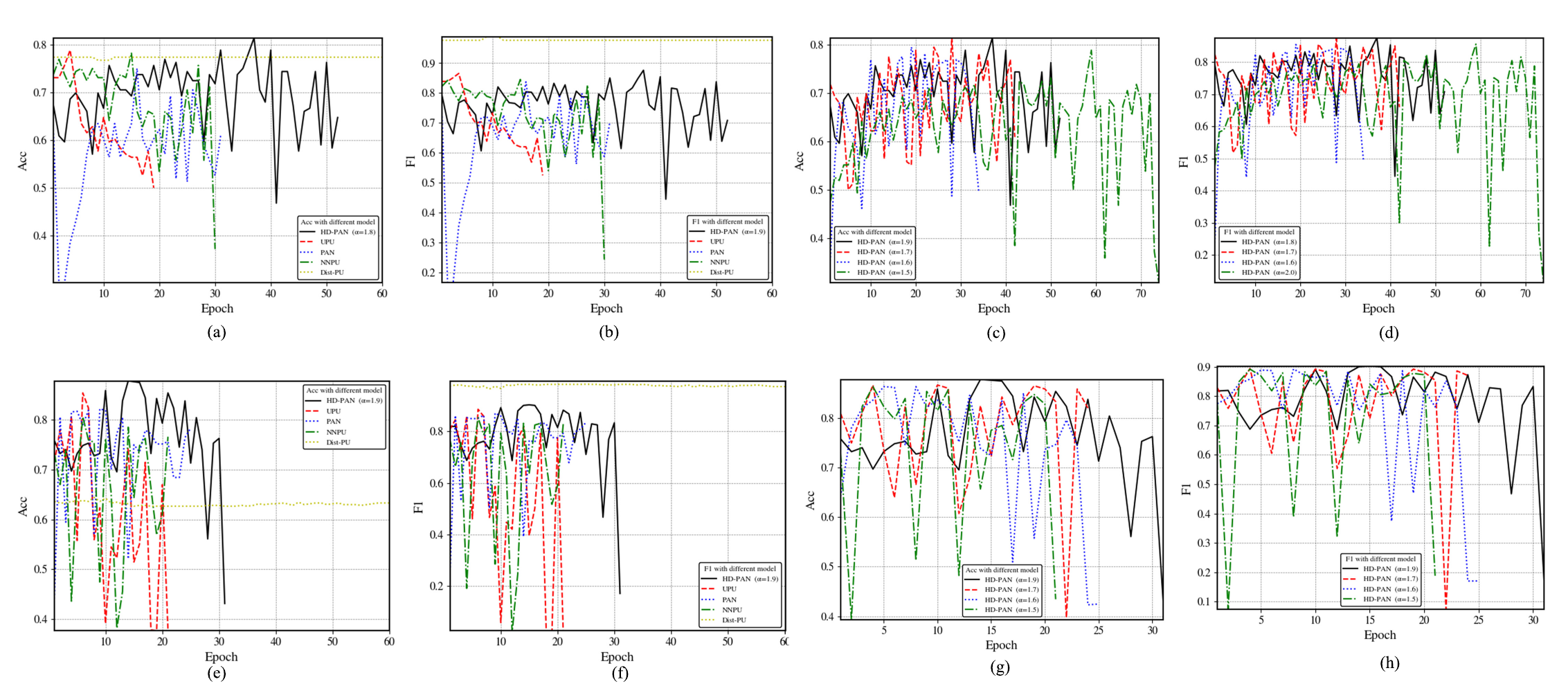}
	
	\caption{Visualization of Quantitative Evaluation: The figure presents a comprehensive quantitative evaluation based on inter-class and intra-class experimental results conducted on two medical image datasets, namely BreastMNIST and PneumoniaMNIST. Subfigures (a) and (b) showcase the inter-class experimental results of BreastMNIST, while (c) and (d) illustrate the intra-class experimental outcomes for the same dataset. Additionally, subfigures (e) and (f) depict the inter-class experimental results for PneumoniaMNIST, and (g) and (h) present the intra-class experimental results for PneumoniaMNIST.}
	\label{fig_2}
\end{figure*}
\begin{figure*}
	\centering
	\includegraphics[width=0.9\linewidth]{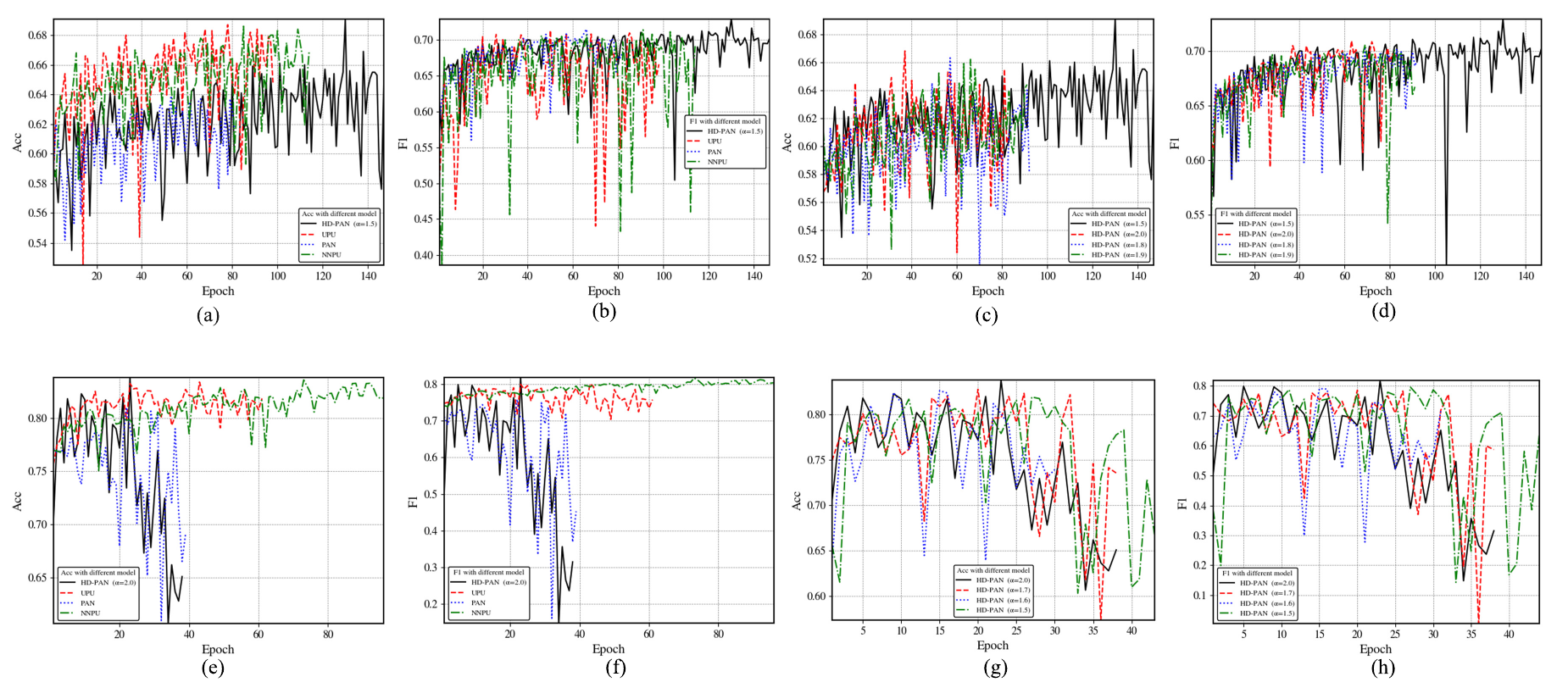}
	
	\caption{Visualization of Quantitative Evaluation: The figure presents a detailed quantitative evaluation through inter-class and intra-class experimental results conducted on two medical image datasets, OCTMNIST and BloodMNIST. Subfigures (a) and (b) illustrate the inter-class experimental outcomes for OCTMNIST, while (c) and (d) showcase the intra-class experimental results for the same dataset. Furthermore, subfigures (e) and (f) depict the inter-class experimental results for BloodMNIST, and (g) and (h) showcase the intra-class experimental outcomes for BloodMNIST, respectively. }
	\label{fig_3}
\end{figure*}
\begin{figure*}
	\centering
	\includegraphics[width=0.9\linewidth]{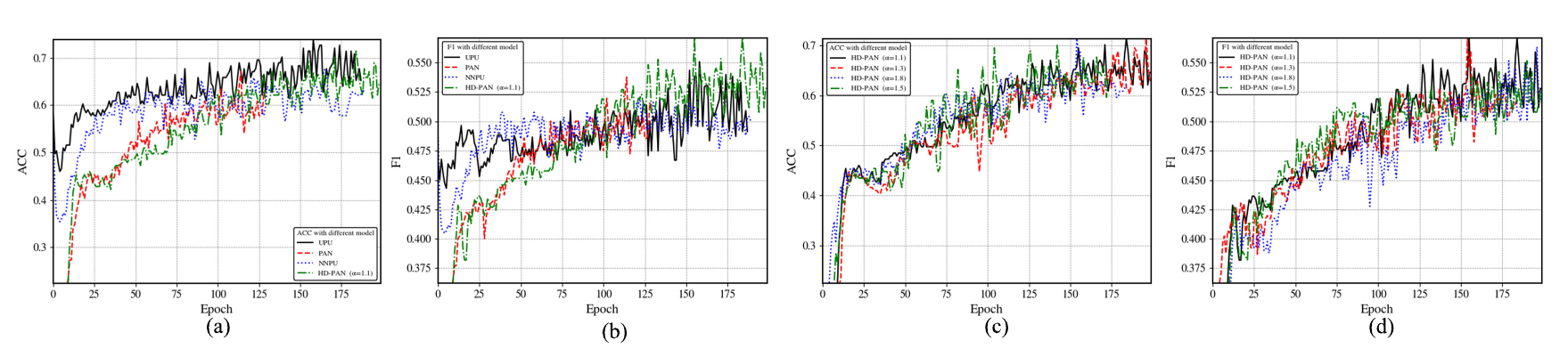}
	
	\caption{Quantitative Evaluation Visualization: The figure illustrates the intra-class and inter-class experimental outcomes for the AMD dataset. Subfigures (a) and (b) depict the inter-class experimental results, while subfigures (c) and (d) showcase the intra-class experimental outcomes.}
	\label{fig_4}
\end{figure*}

\subsubsection{Compare Experimental Models}	
The comparative experimental models employ in this study are NPU \cite{p21}, UUPU \cite{p17}, PAN \cite{p20}, and Dist-PU \cite{p39}. Each model brings unique contributions to the field of PU learning, as outlined below:
\begin{itemize}
	\item UPU (Du Plessis, 2015) \cite{p21}: UPU introduces a convex formulation with double hinge loss for PU learning, effectively addressing bias issues. The authors not only prove the convergence rate of the estimators but also demonstrate the computational efficiency of their non-convex method in experimental settings.
	
	\item NNPU (Ryuichi Kiryo, 2017) \cite{p17}: NNPU proposes a non-negative risk estimator for PU learning, offering robustness against overfitting. This model allows for greater flexibility in utilizing limited positive data, making it advantageous in scenarios where positive data is scarce.
	
	\item PAN (Wenpeng Hu, 2019) \cite{p20}: PAN leverages adversarial training to accomplish PU learning classification tasks. By avoiding the need for class prior probabilities, PAN exhibits superior performance compared to UPU and NNPU, particularly in real-world operations where class prior probabilities are often unavailable.
	
	\item Dist-PU (Yunrui Zhao, 2022) \cite{p39}: Dist-PU proposes a model that emphasizes the consistency of label distributions between predicted and true labels. This approach enhances the accuracy and reliability of PU learning by incorporating the principle of label distribution consistency.
\end{itemize}

It is important to note that both UPU and NNPU rely on class prior probabilities as inputs, which may not be readily available in real-world scenarios. In our experiments, we provide the correct class prior probabilities in advance. Despite this favorable condition, UPU and NNPU still exhibit inferior performance compare to PAN, which does not rely on class prior probability inputs. For consistency, we adopt the same network structure, parameters, and optimization methods as the original models when using UPU and NNPU.
\begin{figure}
	\centering
	\includegraphics[width=0.7\linewidth]{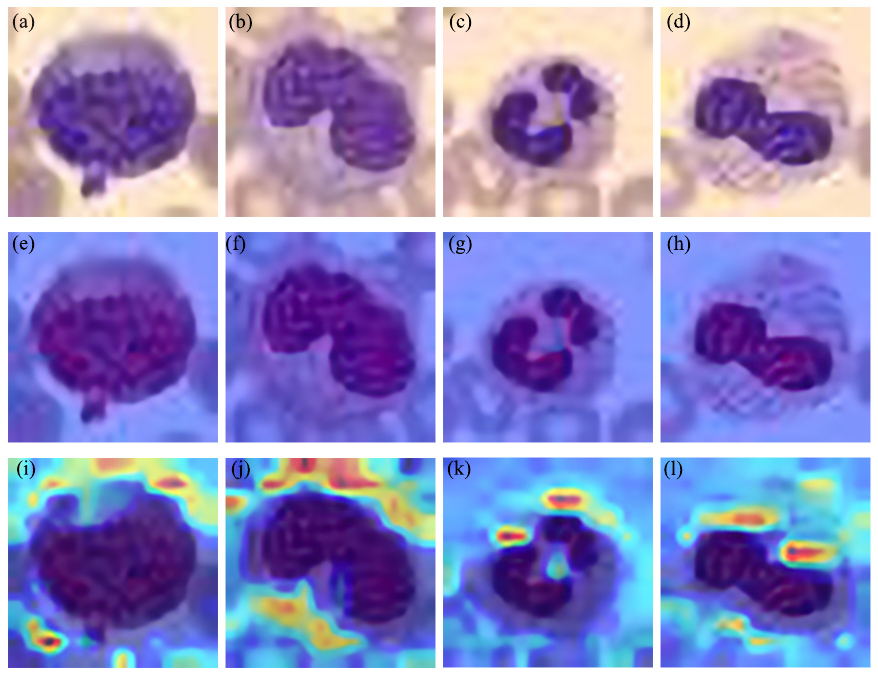}
	\caption{The figure presents a comparison of attention maps between PAN (second row) and HD-PAN (third row), with the first row depicting the original images. }
	\label{fig_4-1}
\end{figure}

\subsubsection{Network Structure and Hyperparameter}
To ensure a fair comparison of experimental results, we adopt the same network architecture as the comparison models to construct our classifier. Specifically, for the BreastMNIST, PneumoniaMNIST, and OCTMNIST datasets, which consist of grayscale images with a size of $28 \times 28$ pixels, we utilize a three-layer Multilayer Perceptron (MLP). As for the BloodMNIST dataset, which comprises RGB-format images with the same size, we employ a Convolutional Neural Network (CNN) for validation. The CNN architecture is defined as follows: $\left( {{\rm{28 }} \times {\rm{ 28 }} \times {\rm{ }}3} \right) - \left[ {C\left( {3{\rm{ }} \times {\rm{ }}3,{\rm{ 84}}} \right)} \right]{\rm{ }} - {\rm{ }}C\left( {3{\rm{ }} \times {\rm{ }}3,{\rm{ 84}},{\rm{ }}2} \right)$ $- {\rm{ }}\left[ {C\left( {1{\rm{ }} \times {\rm{ }}1,{\rm{ 168}}} \right)} \right]$ $ - {\rm{ }}C\left( {1{\rm{ }} \times {\rm{ }}1,{\rm{ 8}}} \right)$ $- {\rm{ }}1000{\rm{ }} - {\rm{ }}1000{\rm{ }} - {\rm{ }}1$. In the architecture, $C\left( {3{\rm{ }} \times {\rm{ }}3,{\rm{ 84}}} \right)$ represents a convolutional layer with a $3 \times 3$ kernel and 84 channels.

Regarding the hyperparameters in PU learning, three key factors include the learning rate, the number of labeled data points, and the batch size. However, in our HD-PAN approach, we construct a novel objective function based on Hölder divergence. This function requires the computation of Hölder exponents $\alpha$ and $\beta$. Since $\alpha$ and $\beta$ are Hölder conjugate exponents, we assign a value to $\alpha$, and $\beta$ is computed accordingly. Moreover, the batch size is set to 64, and we experiment with learning rates of 0.4, 0.5, 0.6, 0.7, and 0.8, respectively. Additionally, we discuss the Hölder exponent $\alpha$ as different divergences, such as Cauchy-Schwarz divergence \cite{p24}, Bregman divergence \cite{p24}, and $\gamma$-divergence \cite{p24}, are formed by different values of $\alpha$ in Hölder divergence. Therefore, it is necessary to investigate the value of $\alpha$, considering values of 1.5, 1.6, 1.7, 1.8, 1.9, and 2.0, respectively.

\subsubsection{Experimental Steps}
During the experiment, we sequentially execute UPU, NNPU, Dist-PU, and PAN on five medical image datasets. Throughout the training process, we modify the parameters of these models to examine the impact of different learning rates on the training outcomes. Additionally, we investigate the influence of the Hölder parameter $\alpha$ on the training results by varying its value during the training process. Subsequently, for each learning rate, we evaluate the optimal training impact.

\subsubsection{Experimental Results and Analysis}
For binary classification tasks, accuracy (Acc) and the $F_1$-score are important metrics. In the context of PU learning, the primary objective is to identify all positive data points through the algorithm for further analysis. If the model overlooks some positive samples while detecting others, its performance will be significantly compromised, failing to meet the requirements of robustness and interpretability. In the realm of medical image-aided diagnosis, misclassifying several diseased samples as disease-free patients can lead to numerous misdiagnoses and missed treatment opportunities, which is undesirable. In PU learning, the emphasis is placed on the identification and classification of positive samples, and the $F_1$-score, as the harmonic mean of precision and recall, provides a comprehensive measure for positive samples. Consequently, when evaluating PU learning algorithms, the $F_1$-score is more suitable as the primary metric.
\renewcommand\arraystretch{1.0}
\begin{table}[t]
	\setlength{\belowdisplayskip}{0pt}
	\setlength{\abovedisplayskip}{0pt}
	\setlength{\abovecaptionskip}{0pt}
	\centering
	\scriptsize
	\caption{Quantitative evaluation based on intra-class experimental results (accuracy and $F_1$-score) for breastMNIST, pneumoniaMNIST, OCTMNIST, bloodMNIST, and AMD datasets.}
	\setlength{\tabcolsep}{12pt}
	\begin{tabular}{p{1.5cm}p{2.3cm}p{1.0cm}p{1.2cm}}  
		\toprule[1.5pt]
		{Dataset}&	Models &	Accuracy &	$F_1$-Score\\
		\midrule[0.5pt]
		{BreastMNIST}&UPU \cite{p21}&	0.7885&	0.8649\\
		&PAN \cite{p20}&	0.7500&	0.8367\\
		&NNPU \cite{p17}&	0.7821&	0.8448\\
		&{Dist-PU}\cite{p39}&	0.7737&	\textcolor{red}{0.9879}\\
		&HD-PAN ($\alpha=1.8$)&	\textcolor{red}{0.8141}&	0.8755\\
		\midrule[0.5pt]
		{PneumoniaMNIST}&UPU \cite{p21}&	0.8542&	0.8853\\
		&PAN \cite{p20}&	0.8221&	0.8703\\
		&NNPU \cite{p17}&	0.8061&	0.8530\\
		&Dist-PU \cite{p39}&	0.6378&	\textcolor{red}{0.9826}\\
		&HD-PAN ($\alpha=1.9$)&	\textcolor{red}{0.8782}&0.9003\\
		\midrule[0.5pt]
		{OCTMNIST}&UPU \cite{p21}&	0.6870&	0.6756\\
		&PAN \cite{p20}&	0.6370&	0.7151\\
		&NNPU \cite{p17}&	0.6860&	0.6990\\
		&HD-PAN ($\alpha=1.5$)&	\textcolor{red}{0.6910}&\textcolor{red}{0.7295}\\
		\midrule[0.5pt]
		{BloodMNIST}&UPU \cite{p21}&	0.8340&	0.8010\\
		&PAN \cite{p20}&	0.8100&	0.7587\\
		&NNPU \cite{p17}&	0.8360&	0.8173\\
		&HD-PAN ($\alpha=2.0$)&	\textcolor{red}{0.8385}&\textcolor{red}{0.8196}\\
		\midrule[0.5pt]
		{AMD}&UPU \cite{p21}&\textcolor{red}{0.7391}&0.5510\\
		&PAN \cite{p20} &0.6708&0.5378\\
		&NNPU \cite{p17}&0.6584&0.5210\\
		&HD-PAN ($\alpha=1.8$)&0.7143&\textcolor{red}{0.5717}\\
		\bottomrule[1.5pt]
	\end{tabular}
	\label{tab02}
\end{table}	
\renewcommand\arraystretch{1.0}
\begin{table}
	\setlength{\belowdisplayskip}{0pt}
	\setlength{\abovedisplayskip}{0pt}
	\setlength{\abovecaptionskip}{0pt}
	\centering
	\scriptsize
	\caption{Quantitative Evaluation of inter-class Experimental Results (Accuracy and $F_1$-Score) on BreastMNIST, PneumoniaMNIST, OCTMNIST, BloodMNIST, and AMD Datasets.}
	\setlength{\tabcolsep}{8pt}
	\begin{tabular}{p{1.75cm}p{3.4cm}p{1.0cm}p{1.2cm}}  
		\toprule[1.5pt] 
		{Dataset}&	Models &	Accuracy &	$F_1$-Score\\
		\midrule[0.5pt]
		{BreastMNIST}&HD-PAN ($\alpha=1.5,lr=0.6$)&	0.7692&	0.8429\\
		&HD-PAN ($\alpha=1.6,lr=0.6$)&	0.7949&	0.8547\\
		&HD-PAN ($\alpha=1.7,lr=0.8$)&	\textcolor{red}{0.8141}&	0.8724\\
		&HD-PAN ($\alpha=1.8,lr=0.7$)&\textcolor{red}{0.8141}&\textcolor{red}{0.8755}\\
		&HD-PAN ($\alpha=1.9,lr=0.4$)&	0.7885&	0.8445\\
		&HD-PAN ($\alpha=2.0,lr=0.4$)&	0.7885&	0.8560\\
		\midrule[0.5pt]
		{PneumoniaMNIST}&HD-PAN ($\alpha=1.5,lr=0.6$)&	0.8638&	0.8922\\
		&HD-PAN ($\alpha=1.6,lr=0.5$)&	0.8638&	0.8927\\
		&HD-PAN ($\alpha=1.7,lr=0.7$)&	0.8670&	0.8920\\
		&HD-PAN ($\alpha=1.8,lr=0.7$)&	0.8510&	0.8774\\
		&HD-PAN ($\alpha=1.9,lr=0.8$)&	\textcolor{red}{0.8782}&\textcolor{red}{0.9003}\\
		&HD-PAN ($\alpha=2.0,lr=0.5$)&	0.8574&	0.8894\\
		\midrule[0.5pt]
		{OCTMNIST}&HD-PAN ($\alpha=1.5,lr=0.4$)&\textcolor{red}{0.6910}&\textcolor{red}{0.7295}\\
		&HD-PAN ($\alpha=1.6,lr=0.6$)&	0.6550&	0.6588\\
		&HD-PAN ($\alpha=1.7,lr=0.7$)&	0.6610&	0.6903\\
		&HD-PAN ($\alpha=1.8,lr=0.6$)&	0.6630&	0.6847\\
		&HD-PAN ($\alpha=1.9,lr=0.4$)&	0.6620&	0.6920\\
		&HD-PAN ($\alpha=2.0,lr=0.4$)&	0.6680&	0.7049\\
		\midrule[0.5pt]
		{BloodMNIST}&HD-PAN ($\alpha=1.5,lr=0.7$)&	0.8194&	0.7962\\
		&HD-PAN ($\alpha=1.6,lr=0.7$)&	0.8264&	0.7913\\
		&HD-PAN ($\alpha=1.7,lr=0.8$)&	0.8278&	0.7861\\
		&HD-PAN ($\alpha=1.8,lr=0.8$)&	0.8185&	0.7761\\
		&HD-PAN ($\alpha=1.9,lr=0.7$)&	0.8173&	0.7870\\
		&HD-PAN ($\alpha=2.0,lr=0.7$)&	\textcolor{red}{0.8385}&	\textcolor{red}{0.8196}\\
		\midrule[0.5pt]
		{AMD}&HD-PAN ($\alpha=1.8,lr=0.8$)&\textcolor{red}{0.7143}&\textcolor{red}{0.5714}\\
		&HD-PAN ($\alpha=1.3,lr=0.6$)&0.7143&0.5636\\
		&HD-PAN ($\alpha=1.5,lr=0.4$)&0.7019&0.5455\\
		&HD-PAN ($\alpha=1.4,lr=0.6$)&0.6894&0.5517\\
		&HD-PAN ($\alpha=1.7,lr=0.6$)&0.6894&0.5517\\
		&HD-PAN ($\alpha=1.1,lr=0.8$)&\textcolor{red}{0.7143}&\textcolor{red}{0.5714}\\
		\bottomrule[1.5pt]
	\end{tabular}
	\label{tab01}
\end{table}
\subsubsection{Performance of HD-PAN on Different Hölder Parameters $\alpha$}
Using the BreastMNIST dataset as an example, we conducted experiments to investigate the effects of different learning rates and Hölder parameters ($\alpha$) on the performance of HD-PAN. The aim is to identify the best combination of parameters based on accuracy and the $F_1$-score. For instance, when $\alpha$ is set to 1.8 and the learning rate to 0.7, HD-PAN achieves the best training results on the BreastMNIST dataset, with an accuracy of 0.8141 and an $F_1$-score of 0.8755.

It is important to note that the termination condition for training was set at 15 epochs, with the criterion that the $F_1$-score should not increase by less than 0.01.

Table \ref{tab01} and Figures \ref{fig_2}--\ref{fig_4} illustrate that different values of $\alpha$ can yield varying optimal training results for the same dataset. Taking the BloodMNIST dataset as an example, when $\alpha$ is set to 2.0 and the learning rate to 0.7, the accuracy rate reached 0.8384, while the $F_1$-score reach 0.8196. However, changing the value of $\alpha$ lead to unsuccessful learning outcomes, regardless of variations in the learning rate. Therefore, adjusting the Hölder parameter $\alpha$ can improve the classification performance of the model for different types of medical imaging datasets.

Following the experiments, we obtain results for the HD-PAN model under various Hölder parameters ($\alpha$) and diverse learning rates (lr). Through a comprehensive comparison of Accuracy and $F_1$-score, we identify the optimal training outcomes and their corresponding learning rates for different Hölder parameters. The training set comprises 240 training samples and 160 validation samples from the AMD, while the test set includes 156 test samples from the same dataset.

To illustrate, let's consider the row corresponding to HD-PAN ($\alpha=1.1, lr=0.8$). In this case, the HD-PAN achieves its peak training performance with a Hölder exponent of $\alpha=1.1$ and a learning rate of 0.8. Under this configuration, the Accuracy reaches 0.7143, and the $F_1$-score is recorded at 0.5714.

\subsubsection{Comparison of HD-PAN and Other PU Learning Models}
This study focuses on the exploration of medical imaging datasets and proposes a novel PU learning algorithm by building upon previous research. The experiment involves the selection of five distinct types of medical imaging datasets, on which different disease classification models are trained. The experimental findings, present in Table \ref{tab02} and Figures \ref{fig_2}--\ref{fig_4}, demonstrate that HD-PAN can achieve optimal classification results for these diverse medical imaging datasets by adjusting its settings. Specifically, HD-PAN surpasses other models in terms of classification accuracy and recall.

In addition to identifying the most suitable model for PU learning, this work also investigates the impact of objective functions based on divergence. By analyzing the experimental results, the performance of HD-PAN is evaluated and compared with other models, providing insights into the effectiveness of divergence-based objective functions in disease classification tasks.

\subsection{Ablation Study}

Attention map play a crucial role in machine learning by providing insights into the neural network's focus areas that contribute to the final categorization decision. In our ablation study, we generate attention maps using two different models and analyze the saliency changes in HD-PAN apply to the BloodMNIST dataset \cite{p25, p26}. This allows us to compare the accuracy and sensitivity of each model in capturing the information distribution within the images. It is important to note that unless specified otherwise, our training setups remain identical except for the ablated alterations. Figure \ref{fig_4-1} illustrates the results obtained from the ablation procedures.

Through our analyses, we find that incorporating HD divergence as a measure significantly enhances the performance of PU learning image classification tasks. The results show minimal improvement when solely utilizing KL divergence within PAN, indicating that the key component of our approach lies in the integration of HD divergence. Additionally, by examining the attention map comparisons presented above, we observe that HD-PAN can more accurately identify the regions of the image that contain crucial information, while PAN's output demonstrates lower sensitivity in this regard.

Overall, our findings underscore the efficacy of HD-PAN and the importance of HD divergence in achieving improved performance in PU learning image classification tasks. Furthermore, the attention map comparisons highlight the superior precision of HD-PAN in pinpointing areas with significant information, distinguishing it from the relatively less sensitive output of PAN.

\subsection{Discussion}
This research focuses on essential semi-supervised classification learning technologies based on positive and unlabeled samples. An adversarial training classification model, HD-PAN, is established using a generative model, and a novel objective function is proposed. The training results on a medical imaging dataset demonstrate that HD-PAN outperforms other common PU learning models in terms of classification accuracy and $F_1$-score. However, PU learning remains an evolving field facing new challenges. Although this work has produced valuable research outcomes, several challenges and directions for further exploration need to be addressed. Based on existing work and ongoing research in this field, several possible study directions are outlined below:

\begin{itemize}
	\item \textbf{Increase in the number of positive samples}: In PU learning, positive samples are typically significantly fewer in number compare to unlabeled samples. However, learning the characteristics of positive samples is crucial for accurately identifying positive and unlabeled data, thus directly impacting classification accuracy. One approach to address this issue is to leverage diffusion models to expand the number of samples based on a limited amount of data. Therefore, combining diffusion models with PU learning can be pursued as a novel research direction.
	
	\item \textbf{Exploration of new objective functions}: Although this work improves the objective function based on KL divergence, it is important to note that this is just one way to measure the gap between two probability distributions. There is still room for further exploration and enhancement in this regard. Moreover, different types of datasets may require different objective functions, potentially influencing the final classification performance. Therefore, it is worth investigating which objective function is more robust for most datasets. Additionally, apart from using divergence formulas, exploring alternative approaches to explain the difference between probability distributions can be a promising new direction for research.
	
	\item \textbf{Investigation of restricted sample size and class imbalance issues in datasets}: In this research, the PU learning classification problem is investigated on several datasets, performing a simple 0-1 classification. However, it is important to acknowledge that different datasets have diverse characteristics, and some datasets, including those analyzed in this research, cannot be classified solely based on image label parity. Hence, it is crucial to thoroughly investigate the specific challenges posed by each dataset and develop a unified PU learning framework that can capture commonalities and characteristics across different types of datasets. By enhancing the model's robustness to various data types, this approach can effectively reduce the costs associated with handling new classification tasks.
	
\end{itemize}

In summary, while this research has made significant advancements in PU learning using positive and unlabeled samples, there remain important avenues for further exploration. The suggested study directions include increasing the number of positive samples through diffusion models, exploring alternative objective functions, and addressing restricted sample size and class imbalance issues in datasets. By delving into these areas, researchers can advance the field of medical image processing and contribute to the development of more effective and robust PU learning models.

\section{Conclusion}
PU learning is a highly prominent research area in machine learning, computer vision, and natural language processing. While advancements in deep learning technology have allowed for the successful resolution of some basic PU learning problems, the expansion of its applicability has given rise to new challenges across various domains. These challenges range from natural language dataset classification to fake review detection, with medical imaging datasets posing increasingly complex problems in disease recognition and classification.

Although researchers have proposed viable alternatives, these methods are not without limitations. To address these limitations, this work proposes an adversarial learning approach based on generative adversarial neural networks. Additionally, two different divergence formulas are introduced, and specific formulations are derived to address diverse scenarios using mathematical principles. This research culminates in the development of a novel PU learning model, HD-PAN, which is thoroughly evaluated and validated using multiple medical imaging datasets.

By leveraging the power of adversarial learning and incorporating specific divergence formulas, this work contributes to advancing PU learning in the field of medical image processing. The proposed HD-PAN demonstrates its efficacy in tackling the intricate challenges associated with disease recognition and classification within medical imaging datasets.

	\ifCLASSOPTIONcaptionsoff
	\newpage
	\fi
	
	\bibliographystyle{IEEEtran}
	\bibliography{IEEEabrv,references.bib}

\begin{thebibliography}{10}
\providecommand{\url}[1]{#1}
\csname url@samestyle\endcsname
\providecommand{\newblock}{\relax}
\providecommand{\bibinfo}[2]{#2}
\providecommand{\BIBentrySTDinterwordspacing}{\spaceskip=0pt\relax}
\providecommand{\BIBentryALTinterwordstretchfactor}{4}
\providecommand{\BIBentryALTinterwordspacing}{\spaceskip=\fontdimen2\font plus
\BIBentryALTinterwordstretchfactor\fontdimen3\font minus \fontdimen4\font\relax}
\providecommand{\BIBforeignlanguage}[2]{{%
\expandafter\ifx\csname l@#1\endcsname\relax
\typeout{** WARNING: IEEEtran.bst: No hyphenation pattern has been}%
\typeout{** loaded for the language `#1'. Using the pattern for}%
\typeout{** the default language instead.}%
\else
\language=\csname l@#1\endcsname
\fi
#2}}
\providecommand{\BIBdecl}{\relax}
\BIBdecl

\bibitem{p1}
C.~Gong, H.~Shi, T.~Liu, C.~Zhang, J.~Yang, and D.~Tao, ``{Loss Decomposition and Centroid Estimation for Positive and Unlabeled Learning},'' \emph{IEEE Transactions on Pattern Analysis and Machine Intelligence}, vol.~43, no.~3, pp. 918--932, 2021.

\bibitem{p2}
F.~Wu and X.~Zhuang, ``{Minimizing Estimated Risks on Unlabeled Data: A New Formulation for Semi-Supervised Medical Image Segmentation},'' \emph{IEEE Transactions on Pattern Analysis and Machine Intelligence}, vol.~45, no.~5, pp. 6021--6036, 2023.

\bibitem{p3}
Y.~Wang, J.~Wang, Z.~Cao, and A.~Barati~Farimani, ``{Molecular Contrastive Learning of Representations via Graph Neural Networks},'' \emph{Nature Machine Intelligence}, vol.~4, no.~3, pp. 279--287, 2022.

\bibitem{p8}
Z.~Zhao, Z.~Zeng, K.~Xu, C.~Chen, and C.~Guan, ``{DSAL: Deeply Supervised Active Learning from Strong and Weak Labelers for Biomedical Image Segmentation},'' \emph{IEEE Journal of Biomedical and Health Informatics}, vol.~25, no.~10, pp. 3744--3751, 2021.

\bibitem{p54}
C.~Gong, Q.~Wang, T.~Liu, B.~Han, J.~You, J.~Yang, and D.~Tao, ``{Instance-Dependent Positive and Unlabeled Learning With Labeling Bias Estimation},'' \emph{IEEE Transactions on Pattern Analysis and Machine Intelligence}, vol.~44, no.~8, pp. 4163--4177, 2022.

\bibitem{p55}
B.~Liu, Z.~Che, H.~Zhong, and Y.~Xiao, ``{A Ranking Based Multi-View Method for Positive and Unlabeled Graph Classification},'' \emph{IEEE Transactions on Knowledge and Data Engineering}, vol.~35, no.~3, pp. 2220--2230, 2023.

\bibitem{p5}
C.~Li, L.~Dong, Q.~Dou, F.~Lin, K.~Zhang, Z.~Feng, W.~Si, X.~Deng, Z.~Deng, and P.-A. Heng, ``{Self-Ensembling Co-Training Framework for Semi-Supervised COVID-19 CT Segmentation},'' \emph{IEEE Journal of Biomedical and Health Informatics}, vol.~25, no.~11, pp. 4140--4151, 2021.

\bibitem{p6}
S.~Zhao, T.~Lau, J.~Luo, I.~Eric, C.~Chang, and Y.~Xu, ``{Unsupervised 3D End-to-End Medical Image Registration with Volume Tweening Network},'' \emph{IEEE Journal of Biomedical and Health Informatics}, vol.~24, no.~5, pp. 1394--1404, 2019.

\bibitem{p7}
P.~Liu and G.~Zheng, ``{Handling Imbalanced Data: Uncertainty-Guided Virtual Adversarial Training With Batch Nuclear-Norm Optimization for Semi-Supervised Medical Image Classification},'' \emph{IEEE Journal of Biomedical and Health Informatics}, vol.~26, no.~7, pp. 2983--2994, 2022.

\bibitem{p80}
F.~Qin, N.~Gao, Y.~Peng, Z.~Wu, S.~Shen, and A.~Grudtsin, ``{Fine-Grained Leukocyte Classification with Deep Residual Learning for Microscopic Images},'' \emph{Computer Methods and Programs in Biomedicine}, vol. 162, pp. 243--252, 2018.

\bibitem{p10}
K.~Han, L.~Liu, Y.~Song, Y.~Liu, C.~Qiu, Y.~Tang, Q.~Teng, and Z.~Liu, ``{An Effective Semi-Supervised Approach for Liver CT Image Segmentation},'' \emph{IEEE Journal of Biomedical and Health Informatics}, vol.~26, no.~8, pp. 3999--4007, 2022.

\bibitem{p11}
T.~Bepler, A.~Morin, M.~Rapp, J.~Brasch, L.~Shapiro, A.~J. Noble, and B.~Berger, ``{Positive-Unlabeled Convolutional Neural Networks for Particle Picking in Cryo-Electron Micrographs},'' \emph{Nature Methods}, vol.~16, no.~11, pp. 1153--1160, 2019.

\bibitem{p14}
X.~Yao, J.~Han, D.~Zhang, and F.~Nie, ``{Revisiting Co-Saliency Detection: A Novel Approach based on Two-Stage Multi-View Spectral Rotation Co-Clustering},'' \emph{IEEE Transactions on Image Processing}, vol.~26, no.~7, pp. 3196--3209, 2017.

\bibitem{p15}
D.~Das and C.~G. Lee, ``{A Two-Stage Approach to Few-Shot Learning for Image Recognition},'' \emph{IEEE Transactions on Image Processing}, vol.~29, pp. 3336--3350, 2019.

\bibitem{p16}
F.~Denis, R.~Gilleron, and M.~Tommasi, ``{Text Classification from Positive and Unlabeled Examples},'' in \emph{Proceedings of the 9th International Conference on Information Processing and Management of Uncertainty in Knowledge-Based Systems, IPMU'02}, 2002, pp. 1927--1934.

\bibitem{p18}
Z.~Xu, Z.~Qi, and J.~Zhang, ``{Learning with Positive and Unlabeled Examples Using Biased Twin Support Vector Machine},'' \emph{Neural Computing and Applications}, vol.~25, pp. 1303--1311, 2014.

\bibitem{p19}
D.~Zhang and W.~S. Lee, ``{A Simple Probabilistic Approach to Learning from Positive and Unlabeled Examples},'' in \emph{Proceedings of the 5th Annual UK Workshop on Computational IPositive-Unlabeled Learning with Non-Negative Risk Estimatorntelligence (UKCI)}, 2005, pp. 83--87.

\bibitem{p17}
R.~Kiryo, G.~Niu, M.~C. Du~Plessis, and M.~Sugiyama, ``{Positive-Unlabeled Learning with Non-negative Risk Estimator},'' \emph{Advances in Neural Information Processing Systems}, vol.~30, 2017.

\bibitem{p20}
W.~Hu, R.~Le, B.~Liu, F.~Ji, J.~Ma, D.~Zhao, and R.~Yan, ``{Predictive Adversarial Learning from Positive and Unlabeled Data},'' in \emph{Thirty-Fifth AAAI Conference on Artificial Intelligence, (AAAI) 2021, Virtual Event, February 2-9, 2021}.\hskip 1em plus 0.5em minus 0.4em\relax AAAI, 2021, pp. 7806--7814.

\bibitem{p21}
M.~Du~Plessis, G.~Niu, and M.~Sugiyama, ``{Convex Formulation for Learning from Positive and Unlabeled Data},'' in \emph{International Conference on Machine Learning}.\hskip 1em plus 0.5em minus 0.4em\relax PMLR, 2015, pp. 1386--1394.

\bibitem{p34}
Y.~Xu, C.~Xu, C.~Xu, and D.~Tao, ``{Multi-Positive and Unlabeled Learning},'' in \emph{IJCAI}, 2017, pp. 3182--3188.

\bibitem{p35}
T.~Gong, G.~Wang, J.~Ye, Z.~Xu, and M.~Lin, ``{Margin based PU Learning},'' in \emph{Proceedings of the AAAI Conference on Artificial Intelligence}, vol.~32, no.~1, 2018.

\bibitem{p36}
J.~Bekker, P.~Robberechts, and J.~Davis, ``{Beyond the Selected Completely at Random Assumption for Learning from Positive and Unlabeled Data},'' in \emph{Machine Learning and Knowledge Discovery in Databases: European Conference, ECML PKDD 2019, W{\"u}rzburg, Germany, September 16--20, 2019, Proceedings, Part II}.\hskip 1em plus 0.5em minus 0.4em\relax Springer, 2020, pp. 71--85.

\bibitem{p37}
H.~Bao, T.~Sakai, I.~Sato, and M.~Sugiyama, ``{Convex Formulation of Multiple Instance Learning from Positive and Unlabeled Bags},'' \emph{Neural Networks}, vol. 105, pp. 132--141, 2018.

\bibitem{p38}
E.~Sansone, F.~G. De~Natale, and Z.-H. Zhou, ``{Efficient Training for Positive Unlabeled Learning},'' \emph{IEEE Transactions on Pattern Analysis and Machine Intelligence}, vol.~41, no.~11, pp. 2584--2598, 2018.

\bibitem{p39}
Y.~Zhao, Q.~Xu, Y.~Jiang, P.~Wen, and Q.~Huang, ``{Dist-PU: Positive-Unlabeled Learning from a Label Distribution Perspective},'' in \emph{Proceedings of the IEEE/CVF Conference on Computer Vision and Pattern Recognition}, 2022, pp. 14\,461--14\,470.

\bibitem{p25}
J.~Yang, R.~Shi, D.~Wei, Z.~Liu, L.~Zhao, B.~Ke, H.~Pfister, and B.~Ni, ``{MedMNIST v2-A Large-Scale Lightweight Benchmark for 2D and 3D Biomedical Image Classification},'' \emph{Scientific Data}, vol.~10, no.~1, p.~41, 2023.

\bibitem{p26}
J.~Yang, R.~Shi, and B.~Ni, ``{MedMNIST Classification Decathlon: A Lightweight AutoML Benchmark for Medical Image Analysis},'' in \emph{IEEE 18th International Symposium on Biomedical Imaging (ISBI)}, 2021, pp. 191--195.

\bibitem{p27}
I.~J. Goodfellow, J.~Pouget{-}Abadie, M.~Mirza, B.~Xu, D.~Warde{-}Farley, S.~Ozair, A.~C. Courville, and Y.~Bengio, ``{Generative Adversarial Nets},'' in \emph{Advances in Neural Information Processing Systems 27: Annual Conference on Neural Information Processing Systems 2014, December 8-13 2014, Montreal, Quebec, Canada}, 2014, pp. 2672--2680.

\bibitem{p28}
I.~Goodfellow, J.~Pouget-Abadie, M.~Mirza, B.~Xu, D.~Warde-Farley, S.~Ozair, A.~Courville, and Y.~Bengio, ``{Generative Adversarial Networks},'' \emph{Communications of the ACM}, vol.~63, no.~11, pp. 139--144, 2020.

\bibitem{p29}
B.~Yu, L.~Zhou, L.~Wang, Y.~Shi, J.~Fripp, and P.~Bourgeat, ``{Sample-Adaptive GANs: Linking Global and Local Mappings for Cross-Modality MR Image Synthesis},'' \emph{IEEE Transactions on Medical Imaging}, vol.~39, no.~7, pp. 2339--2350, 2020.

\bibitem{p30}
U.~Upadhyay, Y.~Chen, T.~Hepp, S.~Gatidis, and Z.~Akata, ``{Uncertainty-Guided Progressive GANs for Medical Image Translation},'' in \emph{Medical Image Computing and Computer Assisted Intervention--MICCAI 2021: 24th International Conference, Strasbourg, France, September 27--October 1, 2021, Proceedings, Part III 24}.\hskip 1em plus 0.5em minus 0.4em\relax Springer, 2021, pp. 614--624.

\bibitem{p31}
M.~Marouf, P.~Machart, V.~Bansal, C.~Kilian, D.~S. Magruder, C.~F. Krebs, and S.~Bonn, ``{Realistic in Silico Generation and Augmentation of Single-Cell RNA-SEQ Data Using Generative Adversarial Networks},'' \emph{Nature Communications}, vol.~11, no.~1, p. 166, 2020.

\bibitem{p32}
C.~Wan and D.~T. Jones, ``{Protein Function Prediction is Improved by Creating Synthetic Feature Samples with Generative Adversarial Networks},'' \emph{Nature Machine Intelligence}, vol.~2, no.~9, pp. 540--550, 2020.

\bibitem{p70}
Y.~Jiang, H.~Chen, M.~H. Loew, and H.~Ko, ``{COVID-19 CT Image Synthesis with a Conditional Generative Adversarial Network},'' \emph{IEEE Journal of Biomedical and Health Informatics}, vol.~25, no.~2, pp. 441--452, 2021.

\bibitem{p71}
J.~Yoon, L.~N. Drumright, and M.~van~der Schaar, ``{Anonymization Through Data Synthesis Using Generative Adversarial Networks (ADS-GAN)},'' \emph{IEEE Journal of Biomedical and Health Informatics}, vol.~24, no.~8, pp. 2378--2388, 2020.

\bibitem{p72}
Y.~Zhao, B.~Ma, P.~Jiang, D.~Zeng, X.~Wang, and S.~Li, ``{Prediction of Alzheimer's Disease Progression with Multi-Information Generative Adversarial Network},'' \emph{IEEE Journal of Biomedical and Health Informatics}, vol.~25, pp. 711--719, 2020.

\bibitem{p73}
B.~Hu, Y.~Tang, E.~I.-C. Chang, Y.~Fan, M.~Lai, and Y.~Xu, ``{Unsupervised Learning for Cell-Level Visual Representation in Histopathology Images With Generative Adversarial Networks},'' \emph{IEEE Journal of Biomedical and Health Informatics}, vol.~23, pp. 1316--1328, 2017.

\bibitem{p74}
J.~Chen, G.~Yang, H.~R. Khan, H.~Zhang, Y.~Zhang, S.~Zhao, R.~H. Mohiaddin, T.~Wong, D.~N. Firmin, and J.~Keegan, ``{JAS-GAN: Generative Adversarial Network based Joint Atrium and Scar Segmentations on Unbalanced Atrial Targets},'' \emph{IEEE Journal of Biomedical and Health Informatics}, vol.~26, pp. 103--114, 2021.

\bibitem{p75}
P.~Sawangjai, M.~Trakulruangroj, C.~Boonnag, M.~Piriyajitakonkij, R.~K. Tripathy, T.~Sudhawiyangkul, and T.~Wilaiprasitporn, ``{EEGANet: Removal of Ocular Artifacts From the EEG Signal Using Generative Adversarial Networks},'' \emph{IEEE Journal of Biomedical and Health Informatics}, vol.~26, pp. 4913--4924, 2021.

\bibitem{p76}
Y.~Zhou, B.~Wang, X.~He, S.~Cui, and L.~Shao, ``{DR-GAN: Conditional Generative Adversarial Network for Fine-Grained Lesion Synthesis on Diabetic Retinopathy Images},'' \emph{IEEE Journal of Biomedical and Health Informatics}, vol.~26, pp. 56--66, 2019.

\bibitem{p77}
Y.~Luo, K.~Chen, L.~Liu, J.~Liu, J.~Mao, G.~Ke, and M.~Sun, ``{Dehaze of Cataractous Retinal Images Using an Unpaired Generative Adversarial Network},'' \emph{IEEE Journal of Biomedical and Health Informatics}, vol.~24, pp. 3374--3383, 2020.

\bibitem{p78}
X.~Yang, Y.~Lin, Z.~Wang, X.~Li, and K.~Cheng, ``{Bi-Modality Medical Image Synthesis Using Semi-Supervised Sequential Generative Adversarial Networks},'' \emph{IEEE Journal of Biomedical and Health Informatics}, vol.~24, pp. 855--865, 2020.

\bibitem{p60}
Z.~Yao, T.~Luo, Y.~Dong, X.~Jia, Y.~Deng, G.~Wu, Y.~Zhu, J.~Zhang, J.~Liu, L.~Yang \emph{et~al.}, ``{Virtual Elastography Ultrasound via Generative Adversarial Network for Breast Cancer Diagnosis},'' \emph{Nature Communications}, vol.~14, no.~1, p. 788, 2023.

\bibitem{p61}
S.~Kench and S.~J. Cooper, ``{Generating Three-Dimensional Structures from a Two-Dimensional Slice with Generative Adversarial Network-based Dimensionality Expansion},'' \emph{Nature Machine Intelligence}, vol.~3, pp. 299 -- 305, 2021.

\bibitem{p62}
A.~G{\"u}emes, C.~S. Vila, and S.~Discetti, ``{Super-Resolution Generative Adversarial Networks of Randomly-Seeded Fields},'' \emph{Nature Machine Intelligence}, vol.~4, pp. 1165--1173, 2022.

\bibitem{p22}
T.~Kanamori, ``{Scale-Invariant Divergences for Density Functions},'' \emph{Entropy}, vol.~16, no.~5, pp. 2611--2628, 2014.

\bibitem{p23}
T.~Kanamori and H.~Fujisawa, ``{Affine Invariant Divergences Associated with Proper Composite Scoring Rules and Their Applications},'' \emph{Bernoulli}, vol.~20, pp. 2278--2304, 2014.

\bibitem{p24}
F.~Nielsen, K.~Sun, and S.~Marchand-Maillet, ``{On H{\"o}lder Projective Divergences},'' \emph{Entropy}, vol.~19, no.~3, p. 122, 2017.

\bibitem{p79}
J.~Zarranz-Ventura, G.~Liew, R.~L. Johnston, W.~Xing, T.~Akerele, M.~McKibbin, L.~Downey, S.~Natha, U.~Chakravarthy, C.~Bailey \emph{et~al.}, ``{The Neovascular Age-Related Macular Degeneration Database: Report 2: Incidence, Management, and Visual Outcomes of Second Treated Eyes},'' \emph{Ophthalmology}, vol. 121, no.~10, pp. 1966--1975, 2014.

\bibitem{p43}
H.~Huang, H.~Xu, X.~Wang, and W.~Silamu, ``{Maximum F1-score Discriminative Training Criterion for Automatic Mispronunciation Detection},'' \emph{IEEE/ACM Transactions on Audio, Speech, and Language Processing}, vol.~23, no.~4, pp. 787--797, 2015.

\end{thebibliography}
\end{document}